\documentclass[12pt,a4paper]{article}

\usepackage{bm}
\usepackage{latexsym}
\usepackage{dcolumn}
\usepackage{amsfonts,amssymb}
\usepackage{graphicx}
\usepackage{epsfig}
\usepackage{psfrag}
\usepackage{nccmath}
\usepackage{moresize}
\usepackage{enumerate}
\usepackage[hang,flushmargin]{footmisc}
\usepackage[titletoc,toc]{appendix}
\usepackage{lipsum}
\usepackage{subcaption}
\usepackage{hyperref}
\usepackage{rotating}
\usepackage{multirow}
\usepackage{multicol}
\usepackage{xfrac}
\usepackage{mathtools}
\usepackage{nccmath}
\usepackage{cite}
\usepackage{tikz}
\usetikzlibrary{positioning}
\usepackage{IEEEtrantools}

\usepackage{stackengine}
\usepackage{scalerel}

\stackMath

\newcommand{\be}{\begin{equation}}
\newcommand{\ee}{\end{equation}}
\newcommand{\bea}{\begin{eqnarray}}
\newcommand{\eea}{\end{eqnarray}}
\newcommand{\bse}{\begin{subequations}}
\newcommand{\ese}{\end{subequations}}
\newcommand{\bce}{\begin{center}}
\newcommand{\ece}{\end{center}}
\newcommand{\bfg}{\begin{figure}}
\newcommand{\efg}{\end{figure}}
\newcommand{\bi}{\begin{itemize}}
\newcommand{\ei}{\end{itemize}}
\newcommand{\bed}{\begin{description}}
\newcommand{\eed}{\end{description}}
\newcommand{\ben}{\begin{enumerate}}
\newcommand{\een}{\end{enumerate}}

\newcommand{\la}{\label}

\newcommand{\fr}{\frac}
\newcommand{\sq}{\sqrt}
\newcommand{\no}{\noindent}

\def\c  {\gamma}

\def\d  {\delta}

\def\f  {\phi}

\def\k  {\kappa}

\def\L  {\Lambda}
\def\m  {\mu}
\def\n  {\nu}

\def\O  {\Omega}

\def\r  {\rho}

\def\s  {\sigma}

\def\vph {\varphi}

\def\le {\left}
\def\ri {\right}

\newcommand{\cM}{\mathcal M}

\newcommand{\cQ}{\mathcal Q}

\newcommand{\nab}{\nabla\!}

\newcommand{\dmt}{\d^{(m)}}

\newcommand{\rmt}{\r^{(m)}}

\newcommand{\rft}{\r^{(\vph)}}
\newcommand{\pft}{p^{(\vph)}}

\newcommand{\Om}{\O^{(m)}}
\newcommand{\Of}{\O^{(\vph)}}
\newcommand{\wf}{w^{(\vph)}}

\newcommand{\sw}{\mathsf w}

\newcommand*\rfra[2]{{}^{\scriptstyle{#1}}\!\!\diagup_{\!\!\scriptstyle{#2}}}

\newcommand{\bdm}{\begin{displaymath}}
\newcommand{\edm}{\end{displaymath}}

\long\def\symbolfootnote[#1]#2{\begingroup%
	\def\thefootnote{\fnsymbol{footnote}}\footnote[#1]{#2}\endgroup}

\begin{document}

\title{\LARGE{Dynamical system analysis of interacting dark energy-matter 
scenarios at the linearized inhomogeneous level}}

\author{Mohit Kumar Sharma\footnote{email: mr.mohit254@gmail.com}~,~ and~
Sourav Sur\footnote{email: sourav@physics.du.ac.in, sourav.sur@gmail.com}
 \\ \\
{\normalsize \em Department of Physics \& Astrophysics}\\
{\normalsize \em University of Delhi, New Delhi - 110 007, India}
}

\date{}

\maketitle

\begin{abstract}
We carry out the dynamical system analysis of interacting dark energy-matter
scenarios by examining the critical points and stability for not just the 
background level cosmological evolution, but at the level of the linear 
density perturbations as well. While an analysis at the background level can 
lead to a stable phase space trajectory implying that the universe eventually 
transpires to a dark energy dominated (de-Sitter) era, a two-fold degeneracy
in the spectrum of the critical points is found to arise in the inhomogeneous 
picture, due to the possible growth and decay of matter density perturbations. 
Analyzing the phase space dynamics of the growth factor, we show that it 
turns out to be greater than unity initially, for one of the critical points,
and leads to a stable configuration as the fluctuations in the matter density 
die out asymptotically. As to the growth index, we show that the only
trajectory which is physically plausible is the one that evolves mildly at 
high redshifts and gets steeper as time progresses. However, such a 
trajectory amounts to the average value of the growth index, throughout 
the expansion history of the universe, not much deviated from the value 
$6/11$, corresponding to the background $\L$CDM cosmology.

\end{abstract}

\section{Introduction}

In a complicated system such as our universe, which passes through various 
stages of evolution and is presently undergoing an accelerated expansion, 
driven supposedly by a dark energy (DE) component
\cite{CST-rev,FTH-rev,AT-book,wols-ed,MCGM-ed}, 
it is natural to ask: {\it will it ever get stabilized or atleast approach 
to a quasi-stable stage asymptotically?} The best way to address this is to
resort to the dynamical analysis that enables one to assert the qualitative 
description of evolving systems without the prior need of any initial 
conditions. As is well-known, observations such as that of the Cosmic 
Microwave Background (CMB)
\cite{wmap9-CP,wmap9-Fin,Planck15-CP,Planck15-DEMG,Planck18-CP}, 
large-scale structures (LSS)
\cite{rsdnev,rsdhou}, 
type-$1$a Supernovae
\cite{bet-SN,scol-SN}, 
and so on, generally concord to the cosmological constant ($\L$) candidature
of the DE, or more specifically the $\L$CDM model (where CDM stands for the 
cold dark matter). However, because of the severe theoretical problems, viz. 
{\em fine-tuning} and {\em coincidence} associated with $\L$ (or the $\L$CDM),
one's focus gets shifted to the dynamical DE, possibly from the scalar fields, 
such as quintessence
\cite{CRS-quin,CLW-quin,tsuj-quin}, 
k-essence
\cite{AMS2000-kess,AMS2001-kess,MCLT-kess,schr-kess,SSSD-kess,SS-dquin},
etc., or from the so-called {\em modified gravity} (MG) scenarios
\cite{NO-mgDE,tsuj-mgDE,CFPS-mg,JBE-mgDE,papa-ed,NOO-mg,SSASB-MST},
that go beyond the realm of General Relativity (GR).

Apart from the extensive studies of the possible consequences of a dynamically
evolving DE, a considerable interest has recently been developed on the 
interaction(s) or even the unification of the DE component with the matter 
field(s), from various perspectives
\cite{WS-intDE,amend-intDE,CPR-intDE,FP-intDE,CW-intDE,CHOP-intDE,CHP2008-intDE,
CHP2009-intDE,amend-rev,ban-intDE,BBM-uDE,BBPP-uDE,GNP-uDE,FFKB-uDE,CDS-MMT,SDC-MMT}. 
Most notable are the MG cosmological scenarios, or the scalar-tensor (ST) 
equivalents thereof
\cite{FM-ST,frni-ST,BP-ST,BEPS-ST,TUMTY-ST,ENO-ST,CHL-ST,BGP-ST,ST-ST,ST-KT,ASBSS-MSTda,ASBSS-MSTpp}, 
which naturally give rise to the DE-matter (DEM) interactions under conformal
transformations. An interaction of such a sort can have a significant effect 
on the cosmic expansion rate as well as on the formation of large-scales 
structures. Hence, the stability analyses of the corresponding cosmological 
solutions, on the physical ground (i.e. against the density perturbations) 
as well as mathematically (i.e. in the phase space), are crucial. 
While the former has been carried out comprehensively, for a typical (and well-motivated) ST equivalent MG scenario, in a preceding work
we focus on performing the latter (i.e. the dynamical analysis) in this 
paper, by resorting to not just the background cosmological level, but to 
that of the linear density perturbations as well. 

The class of ST theories under consideration is the one with a specific 
(quadratic) form of the non-minimal coupling between a scalar field $\f$ and 
the Ricci scalar curvature $R$, with the potential for $\f$ being just in
the form of a mass $m$, in the original Jordan frame. The corresponding
Lagrangian has its equivalence with that of a wide range of MG scenarios, 
starting from some variants of $f(R,\f)$ to metric-scalar-torsion (MST)
formulations, and so on.
A conformal transformation to the Einstein frame, followed by a field 
redefinition $\f \to \vph$, essentially leads to an exponential coupling of 
$\vph$ with matter (including the CDM), and hence an effective DEM interaction, 
once $\vph$ is considered to be the entity that induces the DE. 

Now, the objective of the dynamical system analysis is to determine the critical 
points (CPs), and examine their nature, or more specifically, the eigenvalues of 
the linear perturbation matrix $\cM$ corresponding to the autonomous set of 
equations, which one suitably constructs out of the governing field equations.
The essence of the CPs, which are the equilibrium solutions of the autonomous
equations, lies in their predictability in asserting whether a system is stable 
or not. Note however that more the number of autonomous variables (which 
characterize the mathematical state of the system), more would be the number
of phase space trajectories representing the plausible ways in which the system 
can evolve. In other words, an increased number of autonomous variables would
inevitably imply a less restrictive dynamical analysis and a better scope of 
figuring out a stable configuration to which the system may transpire. In this 
sense it is much desirable for one to be open to the possibility of the DEM 
interactions, and also take the cosmological density perturbations in 
consideration, so as to have additional autonomous variable(s) enhancing the
dimensionality of the phase space.
\cite{da-BLPS,da-land}. 

As is well known, the LSS can more or less be described correctly just by the
matter density perturbations. The scalar field perturbations are generally too 
small to make any significant contribution, at least at the sub-horizon scales
\cite{koiv-grow,PA-grow,PG-grow,GP-grow}. 
In presence of DEM interaction, the field and matter density perturbations get 
entangled with each other, such that the overall study of the formation of 
LSS requires a full-fledged analysis of each perturbed component. Nevertheless, 
being small in magnitude, one can at least incorporate the on-an-average 
contribution of field perturbations on the matter density perturbations. Larger 
is the coupling between the scalar field and matter, larger would be the field 
contribution on the matter density perturbations. Hence, in view of the high 
dominance of the latter at the sub-horizon scales, it become reasonable to 
consider it (or more precisely, the matter density growth factor $f$) as a 
autonomous variable for our dynamical analysis.
Suitably defining the other dimensionless autonomous variables, we set up 
the full set of autonomous equations, considering for simplicity the overall 
(visible + dark) matter content of the universe to be {\it dust}-like, albeit 
with the interaction with the scalar field $\vph$. 

With the matter density perturbations taken into consideration, the system 
is found to possess more CPs than in the case in which the analysis is 
restricted to the background level cosmology. We qualitatively distinguish
the CPs by examining the eigenvalues of the linear perturbation matrix $\cM$
for the autonomous set of equations, and solve the latter simultaneously to
obtain the phase space trajectories. In addition to having the matter 
perturbation growth factor $f$ as a dynamical variable, we revert back to 
the background level physical variables, viz. the matter density parameter 
$\Om$ and the DEM interaction parameter (or the $\vph$-matter coupling 
parameter) $n$. This then enables us to obtain the $\Om\,$--$\,f$ phase 
portraits, for $n = 0$ ($\L$CDM), and for a certain fiducial setting $n = 
0.1$. 

A similar procedure can be followed while the growth factor is formulated in
terms of the growth index $\c$. However, for the interacting system one finds
an inconsistency with the parametrization $f=[\Om]^\c$ commonly used in the
literature
\cite{WYF-grwpara,PAB-grwpara,BBS-grwpara,BP-grwindx,SBM-grwindx,bat-grwindx,
MBMDR-grwindx,PSG-grwindx,BA-grwindx,HKV-modsel,SS-MSTintDE}.
So, to cope with this, we use an alternative parametrization which we have 
proposed in the earlier paper 
\cite{MS-SS-IJMPD}.
From the corresponding phase portraits, we find that there can be only one
physically plausible trajectory which exhibits a very mild time-evolution
of $\c$ at high redshifts, which picks up gradually near the present epoch
and beyond. For this trajectory, the growth index is not much deviated from 
the $\L$CDM value $6/11$, throughout the expansion history of the universe, 
which implies that for whatever interacting picture we resort to, the
overall (and observationally favoured) $\L$CDM cosmology is not much 
distorted.

\vspace{5pt}
\no 
{\large \sl Conventions and Notations}: Throughout this paper, we use metric 
signature $\, (-,+,+,+)$ and natural units, with the speed of light $c = 1$.
We denote and the gravitational coupling factor by $\, \k = \sq{8 \pi G_N}$, 
where $G_N$ is the Newton's constant, the metric determinant by $g$, and the 
values of parameters or functions at the present epoch by an affixed subscript 
`$0$'. 

\section{Interacting DE-matter framework}
\la{sec:intDEM}

Let us begin with a recourse to the DEM interaction(s) in the standard
Friedmann-Robertson-Walker (FRW) cosmological framework, limiting our 
attention to the scenarios in which such interactions arise naturally. 
Note that the very formulation of a gravitational theory, such as GR, 
leaves the scope of such interactions. To be specific, the conservation
of the total energy-momentum tensor, does not insinuate to that for each
individual component in a multi-component system, such as the universe,
composed of baryons, CDM, DE, radiation etc. 
Since the late-time evolution of the universe is effectively driven by 
two main constituents, viz. the DE and the (visible + dark) matter, any 
breach of their individual self-conservation (due to interactions) can 
affect the dynamical stability of the universe. It is then natural to ask: 
{\em what happens to such an interacting system if one slightly perturbs it 
in the space of the autonomous variables about the critical points?}

Consider now a DE component, induced by a scalar field $\vph$, is interacting 
with matter. Let $T_{\m\n}^{(m)}$ and $T_{\m\n}^{(\vph)}$ denote the respective 
energy-momentum tensors. One has
\be \label{cons-eqn}
\nab^{\,\m} \le[T_{\m\n}^{(m)} + T_{\m\n}^{(\vph)}\ri] = 0 \, . 
\ee 
This in general implies 
\be \label{cons-eqn1}
\nab^{\,\m} \, T_{\m\n}^{(m)} =\, -\, \nab^{\,\m} \, T_{\m\n}^{(\vph)}
=\, \cQ_{\n} \, , 
\ee
where $\cQ_{\n}$ is an interaction four-vector. 

Let us note that an interaction of type $\cQ_{\n}$ although seems to be 
arbitrary in Eq. (\ref{cons-eqn}), arises naturally in the ST equivalent
formulations of MG theories. In other words, the non-minimally coupling
of a scalar field with the Ricci scalar $R$ in the Jordan frame induces 
a coupling between scalar field and matter in the Einstein frame. In this 
paper, we consider a class of such ST formulation for $\cQ_{\n}$ takes 
the form
\cite{MS-SS-IJMPD}:
\be \la{Q}
\cQ_{\n} = \k n \rmt \vph_{,\m} \, , \quad \mbox{such that} 
\quad \k \f = e^{n\k\vph} \, ,
\ee
where $\k^2=8\pi G$, $n$ is a coupling parameter and $\rmt$ is the matter 
density.

For a background cosmological evolution described by the spatially flat 
metric, and driven by a DEM interaction given by Eq. (\ref{Q}), the 
corresponding Friedman and the Klein-Gordon equation for $\vph$ are given as
\bea
H^2 \equiv \le(\fr{\dot{a}}{a}\ri)^2 = \fr{\k^2} 3 \le[ \rmt +\rft \ri] \, , \la{Friedmann} \\
\ddot{\vph} + 3H\dot{\vph} + U_{,\vph} = \fr{3n}{\k} H^2 \Om \dot{\vph} \, ,
\la{KG-eqn}
\eea
where an overdot represents derivative with respect to the cosmic time $t$, 
$H$ is the Hubble parameter, $\rft$ is the energy-density of field $\vph$ and 
$\Om:= \k^2\rmt/(3H^2)$ is the matter density parameter. Note that for our 
analysis we consider an exponential potential of the form 
$U(\vph)=\L e^{-2\k n\vph}$ (where $\L$ is the value of $U$ at present epoch 
$t=t_0$), which lead to exact background solutions, as pointed out in ref. 
\cite{MS-SS-IJMPD}.

Now, we know that a given matter density perturbation $\d \rmt$ plays the
all-important role in the LSS formation. However, in presence of an interaction
between matter and the DE induced by a field $\vph$, a small but not necessarily
negligible effect of $\vph$ on the LSS may result. This may be perceived from 
the equations of motion of the matter density contrast $\dmt:= \d \rmt/\rmt$ 
given by 
\cite{MS-SS-IJMPD},
\be \la{ddmt}
\dmt_{,NN}  +\le[2(1-2n^2)-\fr{3 \Om}2 \ri] \dmt_{,N} = \fr{3(1+2n^2)}2 \Om \dmt \, ,
\ee
where $N(t)= \ln a(t)$ is the number of e-foldings. Note that the rightmost 
term, which is nothing but the source term for $\dmt$ gets enhanced by a 
factor of $2n^2$, regardless of its contribution in the drag-force due to 
the background expansion (middle term). Let us emphasize that this effect is 
not just crucial from the perspective of parametric estimations but also from 
the dynamical behavior of the system even up to the perturbative level, as  
more solutions are expected than in the case when the effects of DE perturbations 
are ignorable.

\section{Critical points and their stability conditions}
\la{sec:CP}

In order to perform the dynamical system analysis, let us define following dimensionless quantities:
\be \la{X,Y,U}
X:=\k \fr{\dot \vph}{\sqrt{6}H} \, , \quad Y:= \k \fr{\sqrt{\L} e^{-\k n \vph }}{\sqrt{3}H} \, , \quad f:=  \fr{\dmt_{,N}}\dmt \, ,
\ee
such that from Eq. (\ref{Friedmann}) it follows that
\be \la{mdp}
\Om = 1-X^2-Y^2 \, .
\ee
The coupled set of first-order differential autonomous equations can be worked out as
\bea
X_{,N}&=&-\fr{1} 2 \le(\sq{6} n-3 X\ri) \le(X^2-Y^2-1\ri) \, ,
\la{dxdn} 
\\
Y_{,N}&=&-\fr{1}2 Y \le(2 \sq{6} n X-3 X^2+3 Y^2-3\ri) \, , 
\la{dydn} \\
f_{,N}&=& -\fr{1} 2 \le[2f^2 -f\le(3(X^2-Y^2)-1+2\sq{6}nX \ri) +3 (1+2n^2)(X^2+Y^2-1)  \ri] \, . 
\la{dudn}
\eea
where we have used Eqs. (\ref{Friedmann}), (\ref{KG-eqn}) and (\ref{ddmt}) and relation $dN=Hdt$. Taking the derivative of Eq. (\ref{mdp}) and using Eqs. (\ref{dxdn}) and (\ref{dydn}), we find that
\be \la{dmpd}
\Om_{,N} = \le(X^2+Y^2-1\ri) \le(\sq{6} n X-3 X^2+3 Y^2\ri)
\ee

The solutions or the critical points (CP) of above set of Eqs. (\ref{dxdn}), (\ref{dydn}) and (\ref{dudn}) can be obtained by simultaneously putting
\be \la{dynam-eqns}
\fr{dX}{dN}=\fr{dY}{dN}=\fr{df}{dN}=0 \, .
\ee
The obtained CPs: ($X_c,Y_c$, $f_c$) are enlisted in table (\ref{tab:CP}). In that table, we have shown all the CPs together with their corresponding field energy density parameter $\Of$ and its equation of state parameter $\wf:=\pft/\rft$. Also, due to the invariance in the form of autonomous Eqs. (\ref{dxdn}), (\ref{dydn}) and (\ref{dudn}) under the exchange of $Y \to -Y$, we represent both positive and negative $Y_c$ in a single CP. In other words, both signs will give identical set of eigenvalues and hence stability condition(s).

In order to obtain the stability condition(s) for CPs given in table (\ref{tab:CP}), one requires to add small perturbations ($|\zeta| \ll 1$) to each one of them and to check whether the system still remains intact or gets deviated from its original configuration. 
Let us consider small perturbations around $X_c$, $Y_c$ and $f_c$:
\be \label{pert-CP}
X=X_c + \d X \, , \quad Y=Y_c+\d Y \, \quad \mbox{and}\quad f=f_c+\d f
\, , 
\ee
such that $\zeta \equiv\{\d X,\d Y,\d f\}$ satisfies the following relation:
\be  \label{pert-CP-eqn}
\fr{d\zeta} {dN} = \mathcal{M}\, \zeta \, ,
\ee
where $\mathcal{M}_{(3\times3)}$ is the Jacobian matrix evaluated at $\{X_c, Y_c, f_c\}$. Note that the sign of each eigenvalue of a matrix $\mathcal{M}$ for a given CP determines the stability of autonomous system of equations. In particular, if all the eigenvalues are negative (positive), then that CP is stable (unstable) or saddle otherwise, provided that each eigenvalue is real-valued and non-zero. In table (\ref{table:Eigenvalues}), we have shown all the obtained eigenvalues together with their stability condition(s). Let us now examine the stability criteria of all CPs one by one:

\begin{table*}
\centering
\renewcommand{\arraystretch}{1.6}
\begin{tabular}{|c|ccc|c|c|c|}
\hline
CP & $X_c$ & $Y_c$ & $f_c$ & $\Of$ & $\wf$ & $\sw$ \\
\hline\hline
(a) & $-1$ & $0$ & $0$ & $1$ & $1$ & $1$ \\
\hline
(b) & $1$ & $0$ & $0$ & $1$ & $1$  & $1$ \\
\hline
(c) & $-1$ & $0$ & $1-\sq{6n}$ & $1$ & $1$  & $1$  \\
\hline
(d) & $1$ & $0$ & $1+\sq{6n}$ & $1$ & $1$  & $1$ \\
\hline
(e) & $\sq{\rfra{3}{2n^2}}$ & $\pm \sq{\rfra{3}{2n^2}-1}$ & $2\!-\!\sq{6n^2-2-\rfra{9}{2n^2}}$ & $\rfra{3}{n^2}-1$ & $ \rfra{n^2}{3-n^2}$  & 
$1$ \\
\hline
(f) & $\sq{\rfra{3}{2n^2}}$ & $\pm \sq{\rfra{3}{2n^2}-1}$ & $2\!+\!\sq{6n^2-2-\rfra{9}{2n^2}}$ & $\rfra{3}{n^2}-1$ &  $ \rfra{n^2}{3-n^2}$  & $1$ \\
\hline
(g) & $\sq{\rfra{2}3}\,n$ & $0$ & $n^2-\rfra{3}2$ & $\rfra{2n^2}3$ & $1$  & $\rfra{2n^2}3$ \\
\hline
(h) & $\sq{\rfra{2}3}\,n$ & $0$ & $1+2n^2$ & $\rfra{2n^2}3$ & $1$  & $\rfra{2n^2}3$ \\
\hline
(i) & $\sq{\rfra{2}3}\,n$ & $\pm\sq{1-\rfra{2n^2}3}$ & $0$ & $1$ & $\rfra{4n^2}3-1$  & $\rfra{4n^2}3-1$ \\
\hline
(j) & $\sq{\rfra{2}3}\,n$ & $\pm\sq{1-\rfra{2n^2}3}$ & $-2(1-2n^2)$ & $1$ & $\rfra{4n^2}3-1$  & $\rfra{4n^2}3-1$  \\
\hline
\hline
 \end{tabular}
 \caption{Critical points $X_c, Y_c$ and $f_c$ with their corresponding DE density and equation of state parameter of field $\vph$ and total equation of state parameter of system.}
 \la{tab:CP}
\end{table*}

\begin{table*}[!htb]
\footnotesize
\centering
\renewcommand{\arraystretch}{2}
\begin{tabular}{|c|ccc|c|c|} 
\hline
& \multicolumn{3}{c|}{} &  \\
CP  & & Eigenvalues  & & CP nature \\
\cline{2-4}
 & $\m_1$ & $\m_2$ & $\m_3$ &   \\
\hline\hline
(a) & $1 - \sq{6}\,n$ & $3+\sq{6}\,n$ & $3+\sq{6}\,n$ &   Unstable or saddle \\
\hline
(b) & $1+\sq{6}\,n$ & $3-\sq{6}\,n$ & $3-\sq{6}\,n$ & Unstable or saddle \\
\hline
(c) & $-1+\sq{6}\,n$ & $3+\sq{6}\,n$ & $3+\sq{6}\,n$  & Stable or unstable or saddle \\
\hline
(d) & $-(1+\sq{6}\,n)$ & $3-\sq{6}\,n$ & $3-\sq{6}\,n$ & Stable or unstable or saddle \\
\hline
(e) & $\sq{24n^2\!-\!8\!-\!\rfra{18}{n^2}}$ & $\sq{\rfra{3}2}\le( 
\rfra{3}n-2n\ri)$ & $-\sq{\rfra{3}2}\le(\rfra{3}n-2n\ri)$ & Saddle \\
\hline
(f) & $-\sq{24n^2\!-\!8\!-\!\rfra{18}{n^2}}$  & $\sq{\rfra{3}2}\le( \rfra{3}n-2n\ri)$ & $-\sq{\rfra{3}2}\le( \rfra{3}n-2n\ri)$ & Saddle \\
\hline
(g) &  $\rfra{5}2+n^2$ & $\rfra{3}2-n^2$ & $-\le(\rfra{3}2-n^2\ri)$ & Saddle \\
\hline
(h) & $-\le(\rfra{5}2+n^2\ri)$ & $\rfra{3}2-n^2$ & $-\le(\rfra{3}2-n^2\ri)$ & Saddle \\
\hline
(i) & $-2(1-2n^2)$ & $-3+2n^2$ & $-3+2n^2$ & 
Stable or unstable or saddle \\
\hline
(j) & $2(1-2n^2)$ & $-3+2n^2$ & $-3+2n^2$ & Stable or saddle \\
\hline
\end{tabular}
\caption{Eigenvalues of critical points of table (\ref{tab:CP}) and their corresponding nature. }
 \la{table:Eigenvalues}
\end{table*}

\bi
{
\item CP(a): It is valid for all real values of $n$. It is 
either unstable or saddle but its nature cannot be determined for 
$n=-\sq{3/2}$ and $\sq{1/6}$. It corresponds to a scenario when the DE 
density parameter $\Of=1$ and the growth of matter perturbations $f=0$. 
But due to the steep-fluid like behavior of DE i.e. $\wf=1$ the cosmic
acceleration is not possible. Hence, it is non-physical.

\item CP(b):  It is valid for all real values of parameter $n$. 
This point is either unstable or saddle but its nature cannot be 
determined for $n=-\sq{1/6}$ and $\sq{3/2}$. It corresponds to a state of the universe when $\Of=1$ and $f=0$. However, due to the steep fluid-like nature of DE and hence absence of cosmic acceleration, it is also not physical. Hence, it does not give physical description of the universe. 

\item CP(c):  It is valid for all real values of parameter $n$. 
This point could be stable, unstable or saddle but its nature cannot 
be determined for $n=-\sq{1/6}$ and $\sq{3/2}$. Although, for small value of $n$ i.e. within the saddle parametric regime, this CP could be considered to lie in the deep-matter dominated era when $f\leq 1$. However, due to dominated DE i.e. $\Of=1$ and $\sw=1$ it is not physically possible. Hence, it does not give physical description 
of the universe. 

\item CP(d): It is valid for all real values of parameter $n$. 
This point could be stable, unstable or saddle but its nature cannot 
be determined for $n=-\sq{1/6}$ and $\sq{3/2}$. Although, it is mathematically stable for $n>\sq{3/2}$ but due to the similar reasons mentioned for CP ($c$) it has no physical relevance.

\item CP(e): It is valid for all values of $n$ except when $n=0$. 
This point is only saddle but its nature cannot be determined for $n= \pm \sq{3/2}$ and $\pm \sq{(1+2\sq{7})/6}$. 
Since $\Of=-1+3/n^2$, therefore, one requires $\sq{3/2}<|n|<\sq{3}$ 
for a physically allowed region $0<\Of < 1$, which is too far from allowed observational constraints. Also, since it lacks the cosmic acceleration due to the stiff nature of DE it does not give 
a physical description of the universe.

\item CP(f): It is valid for all values of $n$ excluding $n=0$. 
This point is only saddle but its nature cannot be determined for $n= \pm \sq{3/2}$ and $\pm \sq{(1+2\sq{7})/6}$. Similar, to the reasons mentioned for CP ($e$), it is also not a physically relevant point.

\item CP(g): It is valid for all values of $n$. It is only a saddle point but its nature cannot be determined for $n= \pm \sq{3/2}$.
Due to its saddle behavior, it may correspond to the growth of matter perturbations, lets say, near the recombination epoch for $n>\sq{3/2}$.
Also, for $n\ll \sq{3/2}$, it corresponds to matter dominated epoch with $\sw \to 0$, but then $f$ becomes negative, which results in decay of matter perturbations. Hence, this point does support growth of matter perturbations. Therefore, it is not a physically relevant point.

\item CP(h): It is valid for all values of $n$. It is only a saddle point but its nature cannot be determined for $n= \pm \sq{3/2}$.
Due to its saddle behavior, it may correspond to the growth of matter perturbations, lets say, near the recombination epoch while also providing a physical description of the background at that epoch i.e. $\sw \ll 1$ for $n<\sq{3/2}$. Interestingly, this point indeed show that $f \leq 1$, which is one of the important feature of an interacting DEM scenario. Therefore, this point is of physical relevance in large redshifts.

\item CP(i): It is valid for all values of $n$. It could be 
stable, unstable or saddle but its nature cannot be determined 
for $n=\pm \sq{3/2}$ and $\pm \sq{1/2}$. For $n \in (-\sq{1/2},\sq{1/2})$ this point actually determines the state of the universe in which DE 
becomes dominated and no evolution in the matter density perturbations can be observed. This point is a late-time attractor. For $n=0$, it corresponds to a perfect de-Sitter universe in the future. This point is of physical relevance in far future.

\item CP(j): It is valid for all values of $n$. It is either 
stable or saddle but its nature cannot be determined for $n=\pm 
\sq{3/2}$ and $\pm \sq{1/2}$. Although, at the background level, it correctly describe the DE dominated universe 
but shows a negative growth of matter perturbations for  $n \in (-\sq{1/2},\sq{1/2})$. This results decay of the matter density contrast which is not possible. Hence, it is not physical.
}
\ei

\subsection{Numerical solutions of state variables }
\la{sub:numerical}

\begin{figure}[!t]
\centering
\begin{subfigure}{.49\textwidth} 
  \centering
  \includegraphics[height=7cm, width=8.5cm]{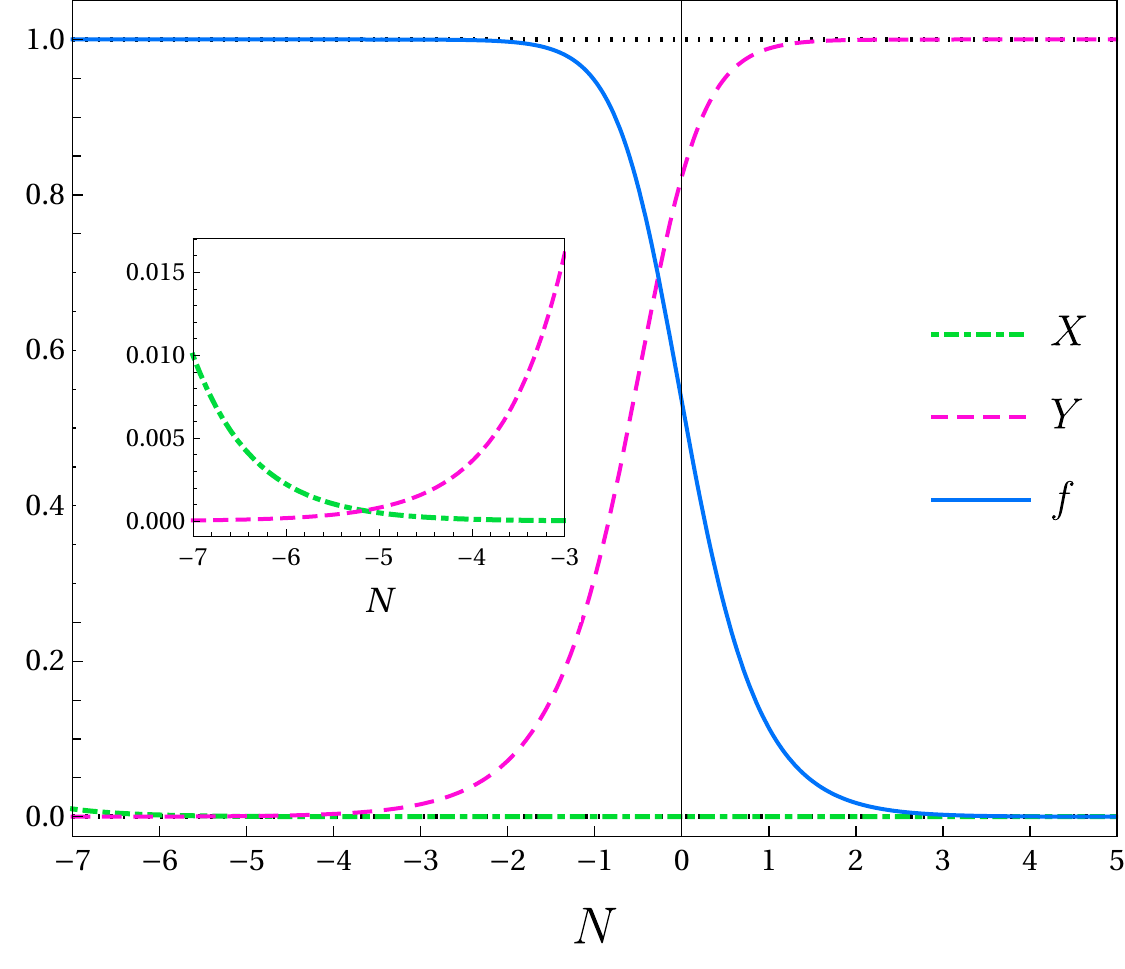}  
  \label{fig:XYU-a}
 \caption{$X(N)$, $Y(N)$ and $f(N)$ for $n=0$.}
\end{subfigure}
\begin{subfigure}{.48\textwidth}
  \centering
  \includegraphics[height=7cm, width=8.5cm]{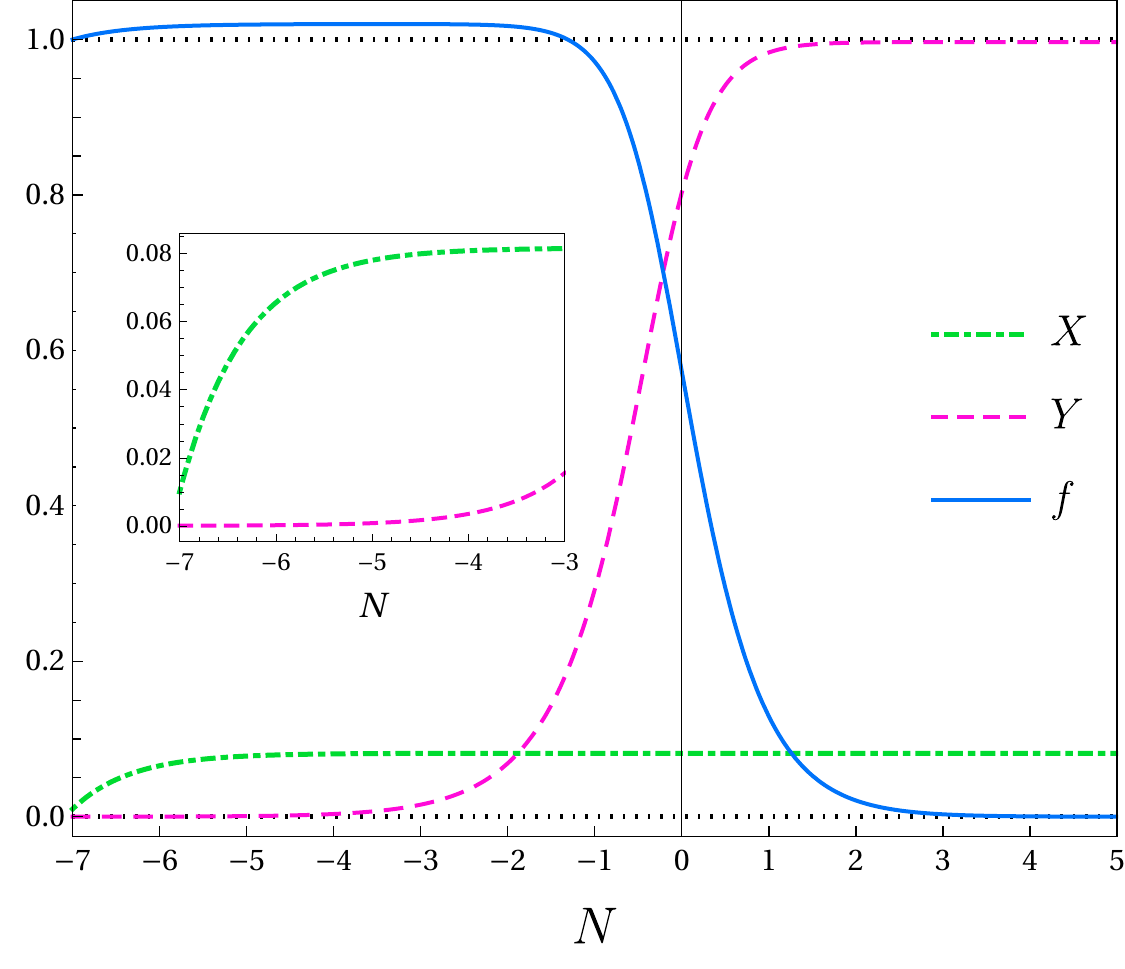}
  \label{fig:XYU-b}  
  \caption{$X(N)$, $Y(N)$ and $f(N)$ for $n=0.1$.}
\end{subfigure}
\caption{\footnotesize Functional profiles of $X$, $Y$ and $f$ for $n=0$ ($\L$CDM) and $0.1$ in the range of $N\in[-7,5]$.}
\label{fig:XYU}
\end{figure}

Now in order to depict the evolutionary behavior of dimensionless variables $X,Y$ and $f$, we numerically solve the coupled system of  Eqs. (\ref{dxdn}), (\ref{dydn}) and (\ref{dudn}). However, before proceeding ahead let us first note that the free parameter $n$ are very tightly constrained by both background as well as at the perturbative level observations. In fact, we have shown in
\cite{MS-SS-IJMPD} 
that from the OHD and RSD-GOLD data, $n \leq 0.1$ upto $1\s$ confidence interval. Therefore, in order to be compatible with observational constraints, we take fiducial values: $n=0$ and $0.1$. Moreover, we take the initial conditions at $N=-7$ or $z=1095$: 
\be
X(-7)=0.01\, , \quad Y(-7)=4 \times 10^{-5} \, , \quad f(-7)=1.
\ee
The obtained evolutionary profiles are shown in Fig. (\ref{fig:XYU}). 
In that figure one observes that $f$ tends to cross the value of unity 
in large enough redshifts. One also finds that variable $X$ is quite small ($\ll 1$) and least evolving variable, whereas, $Y$ increases monotonically. 

Here, let us emphasize that although in fig. (\ref{fig:XYU}) it seems 
that $X$ is constant, but in general it is not true. In fact, we have explicitly shown in insets that for any 
given initial conditions, $X$ always tends to settle to a value of the order of $n$. In other words, changing or perturbing its initial conditions 
does not affect its evolutionary profile. Hence, no matter what initial condition for $X$ one will start with it always finds it near(or equal to) the magnitude of $n$ for any given epoch. It is also interesting to note that $X_c$ for this stable CP($i$) agrees with our previous obtained background solution of $\vph(N)$ 
\cite{MS-SS-IJMPD} 
i.e., 
\be \la{vph_sol}
\vph(N)=\fr{2 n N}\k \, , \quad \mbox{which implies,} \quad X = X_c=\sq{\fr{2}3}n \, .
\ee
Hence, it confirms that this unique solution is both stable physically and dynamically of combined background and perturbative system.

\subsection{Phase space dynamics of matter perturbations}
\la{sub:phaseplane}

In order to depict the phase space dynamics of above system comprised of $X,Y$ and $f$ variables, it is suitable to replace $X,Y$ and their derivatives by $\Om$ and $\Om_{,N}$, respectively, by using Eq. (\ref{mdp}) and (\ref{dmpd}). In this way, not just our phase space dynamics will get reduced to two-dimensions but it will also be more reasonable to have a phase space dynamics between two physically observable quantities i.e., $\Om$ and $f$.
Using above solution (\ref{vph_sol}) together with Eq. (\ref{dxdn}), 
(\ref{dydn}) and (\ref{dudn}), one obtains a set of equations:
\bea \la{dodn-sol}
\Om_{,N} &=& -3 \le( 1 -\fr{2}3 n^2 - \Om \ri) \Om \, , \\
f_{,N} &=& \fr{3}2 \Om \le(1+2 n^2+f\ri)-f \le(2-4 n^2+f\ri)\, .
\la{dudn-sol}
\eea
\begin{figure}[!t]
\centering
\begin{subfigure}{.49\textwidth} 
  \centering
  \includegraphics[height=7cm, width=8.5cm]{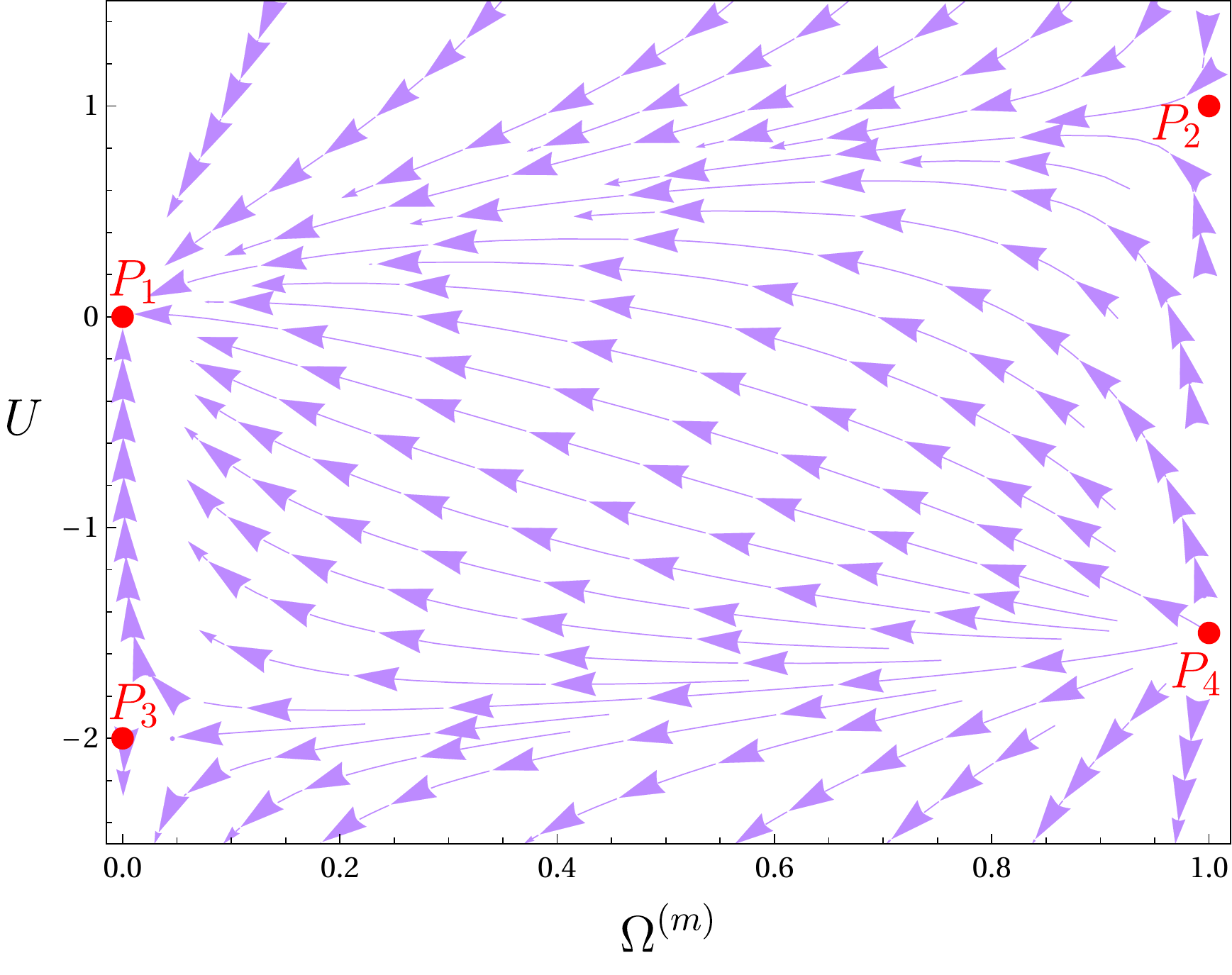}  
  \label{fig:OmU-a}
 \caption{Phase portrait for $n=0$.}
\end{subfigure}
\begin{subfigure}{.48\textwidth}
  \centering
  \includegraphics[height=7cm, width=8.5cm]{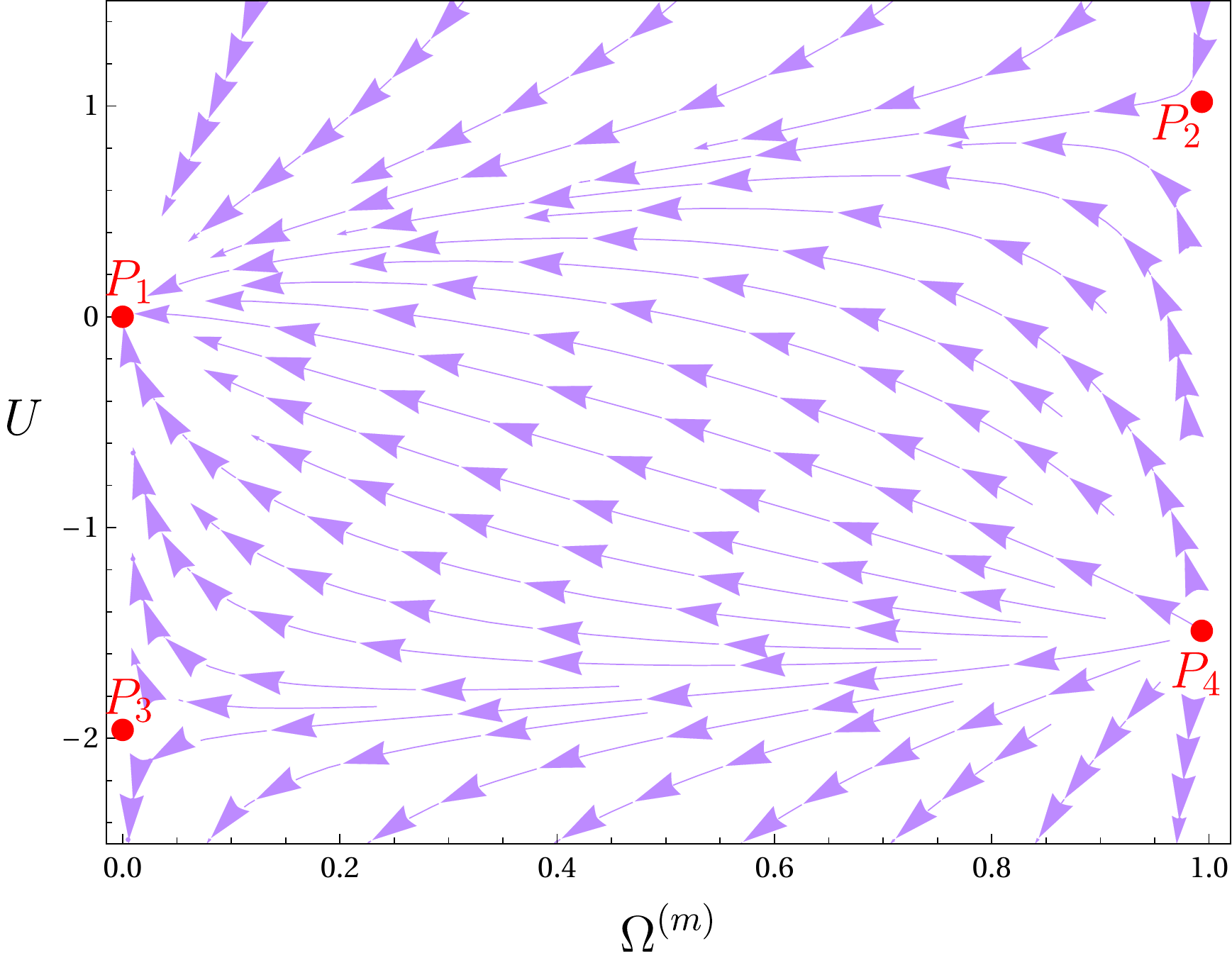}
  \label{fig:OmU-b}  
  \caption{Phase portrait for $n=0.1$.}
\end{subfigure}
\caption{\footnotesize Phase portrait in the space of the matter density 
parameter and growth factor for $n=0$ ($\L$CDM) and $0.1$.}
\label{fig:OmU}
\end{figure}
Solving them simultaneously by putting $\Om_{,N}=f_{,N}=0$, we obtain the phase space dynamics between matter density parameter $\Om$ and growth factor $f$. 
In fig. (\ref{fig:OmU}), we show the phase portrait diagram for $\L$CDM and $n=0.1$. In this figure, the nature of CPs are given as: ($i$) CP($P_1$) corresponds to the stable de-Sitter solution, ($ii$) : CP($P_2$) is a saddle point which may belongs to the deep matter-dominated era, ($iii$) : CP($P_3$) is a saddle, and ($iv$) : CP($P_4$) is an unstable point. Since, both $P_3$ and $P_4$ corresponds to the negative growth of matter perturbations and hence the matter density contrast will behave as: $\dmt\ \propto 
e^{-c N}$ (where $c$ is some positive integration constant), the LSS formation will not occur. Hence, they are physically irrelevant. Therefore, one only expect those trajectories that starts with $P_2$ and ends at $P_1$. In fact, these are the only trajectories which makes the universe to evolve in such a way that growth of matter perturbations declines as matter density parameter decreases with time.

After depicting the possible evolutionary profile of $f$ with $\Om$, it is also necessary to figure out how does the evolutionary profile of growth index ($\c$) follows. Let us recall that, in literature, the growth index is related to the growth factor $f$ via the relation: $f =[\Om] ^\c$. In general, $\c$ is a time-varied function and depends on the background parameters such as $\Om$ and coupling parameters ($n$ in our case). As it can be noticed from fig. (\ref{fig:XYU}) that $f$ crosses the upper barrier of unity in presence of $n$, which is not possible in the above parametrization of $f$ as it is only limited to the minimally coupled scalar field scenarios, such as quintessence. In order to incorporate this feature, the above parametrization of $f$ requires modification. In fact, in 
\cite{MS-SS-IJMPD}, 
we have shown that a suitable modification in the parametrization of $f$ can take the following form:
\be \la{U-param}
f(N) = (1+2n^2)\le[\Om\ri]^{\c(N)} \, .
\ee
Using this in the Eq. (\ref{dudn-sol}), we obtain a first-order differential equation for $\c(N)$:
\be \la{dgammadn}
\c_{,N} = -\frac{4-3 (\Om)^{1-\c}+2 (\Om)^\c+6 \c  (\Om -1)+4 n^2
   \le[(\Om) ^\c +\c -2\ri]-3 \Om}{2 \log (\Om)} \, .
\ee
\begin{figure}[!t]
\centering
\begin{subfigure}{.49\textwidth} 
  \centering
  \includegraphics[height=7cm, width=8.5cm]{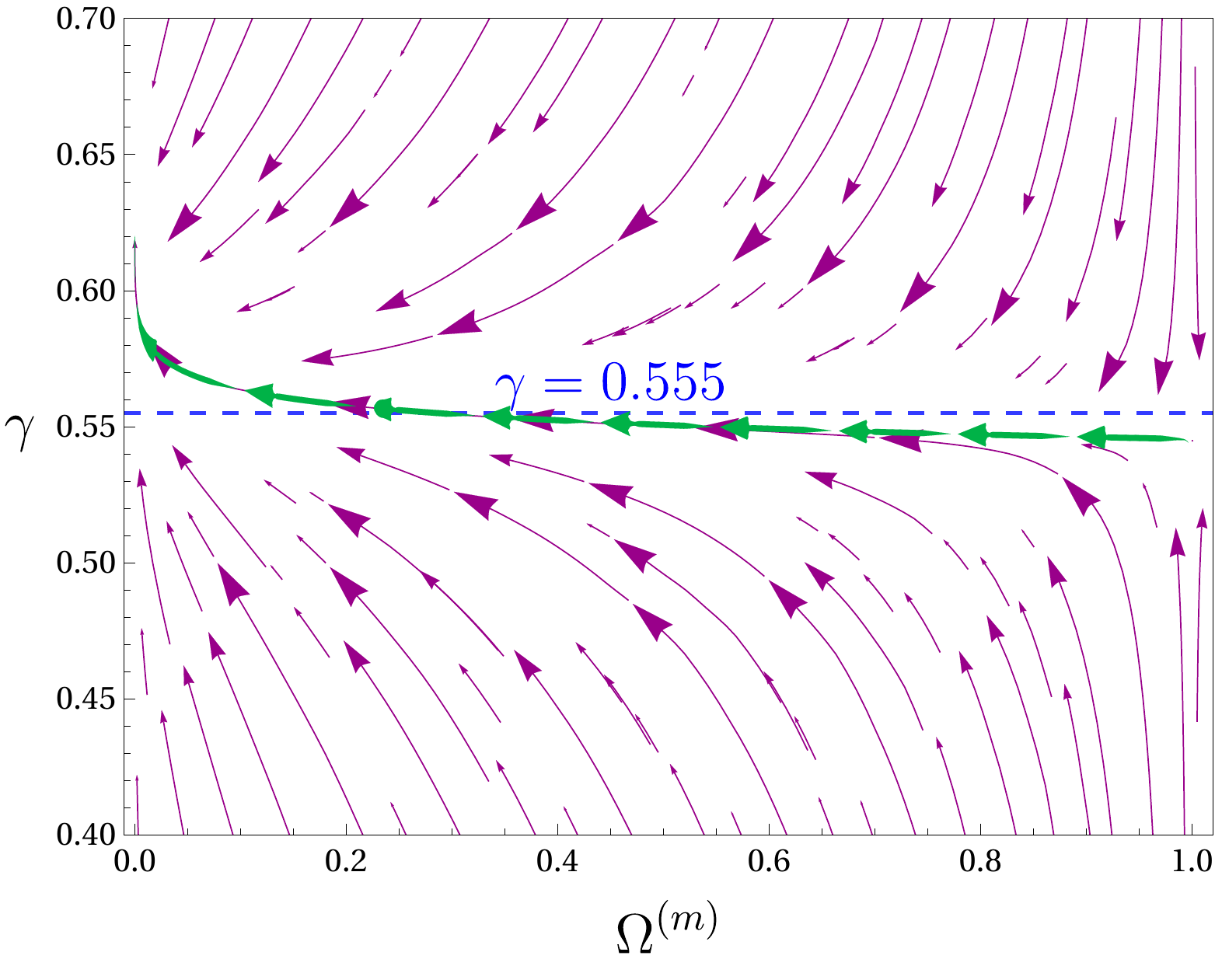}  
  \label{fig:Omgamma-a}
 \caption{Phase portrait for $n=0$.}
\end{subfigure}
\begin{subfigure}{.48\textwidth}
  \centering
  \includegraphics[height=7cm, width=8.5cm]{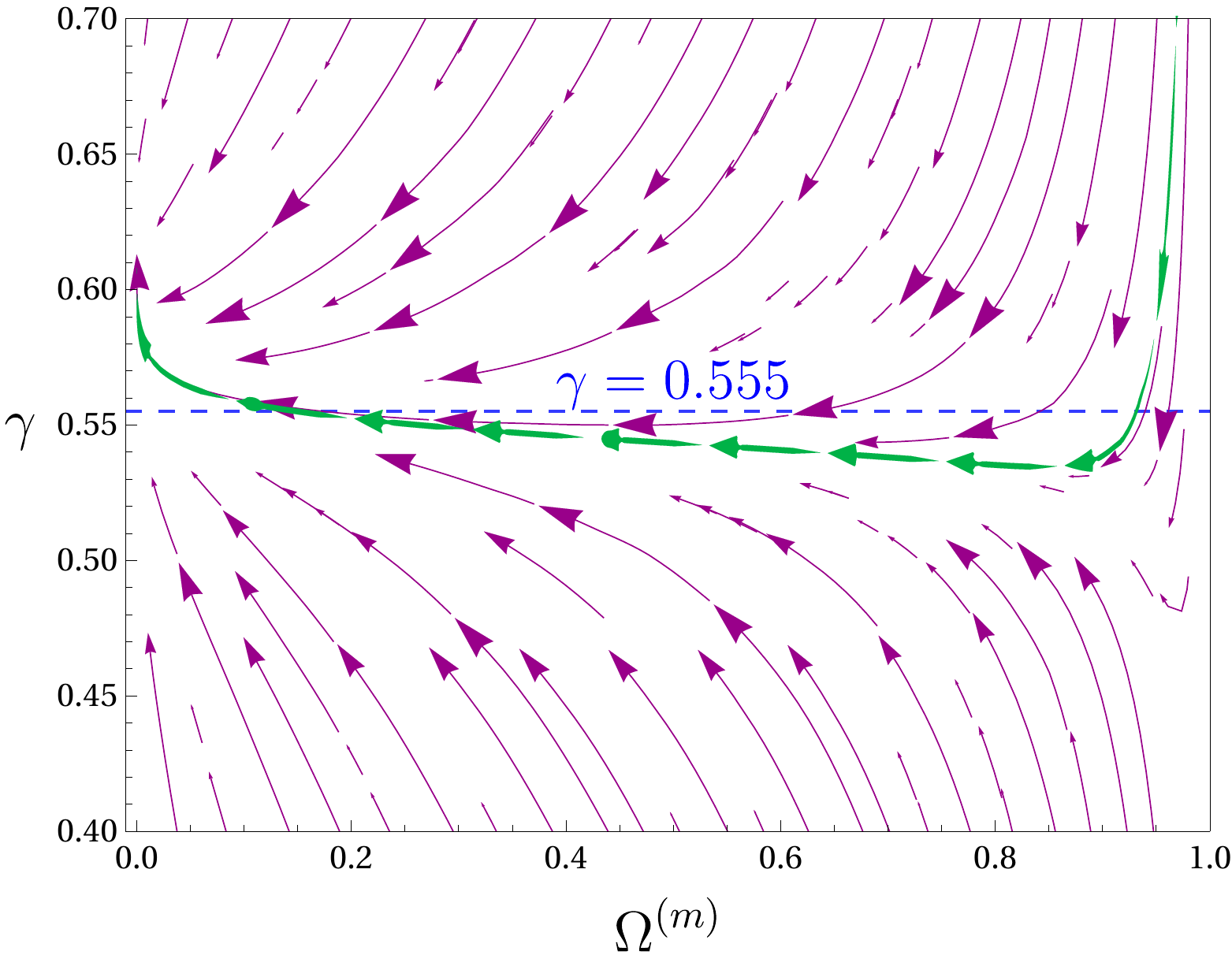}
  \label{fig:Omgamma-b}  
  \caption{Phase portrait for $n=0.1$.}
\end{subfigure}
\caption{\footnotesize Phase portrait in the space of the matter density 
parameter and growth index for $n=0$ ($\L$CDM) and $0.1$.}
\label{fig:Omgamma}
\end{figure}
By putting $\c_{,N}=\Om_{,N}=0$, and then solving Eqs. (\ref{dodn-sol}) and 
(\ref{dgammadn}) simultaneously we obtain the phase space diagram between 
$\Om$ and $\c$ (shown in fig. (\ref{fig:Omgamma})) for both $\L$CDM and 
$n=0.1$. In that 
figure, the green vector-field (thick) is the only solution as it remains 
finite over the entire range of $\Om \in [0,1]$. All other solutions gets 
diverge which either implies extensive amount of growth of matter 
perturbations (when $\c\ll 0.555$) or no structure formation at all (when 
$\c\gg 0.555$). This solution, thus, giving rise to consistent matter 
perturbations growth at the present epoch is also consistent with the 
theoretically calculated value $\c=6/11$ in the deep matter-dominated epoch. 
It is interesting to note that this behavior of increasing $\c$ with $N$ 
agrees with our fitting function of $\c$ in previous paper 
\cite{MS-SS-IJMPD}, 
which also depicts sharp increase in the value of $\c$ in future by remaining mildly dynamic atleast up to the present epoch.

\section{Conclusion}
\la{sec:conclusion}

We have performed a thorough dynamical analysis considering both the cosmological background as well as linear perturbative effects in order to find out the stability of the entire system. The system under consideration involves an interaction between DE and matter which has generated as a consequence of a non-minimal coupling (exponential in nature) between scalar field and matter sector in the Einstein frame.  Since this type of coupling can be obtained in various sort of modified gravity theories such as $f(\f,R)$, metric-scalar-torsion theories, etc., which although have different theoretical motivations but due to the degeneracy at the solution level, they land up to an identical dynamical system. Therefore, it becomes important to perform a dynamical system not just considering the cosmological background but also the linear perturbative level which can incorporate a wide variety of modified gravity theories. Keeping that as an objective we have carried out our dynamical analysis in this paper.

In presence of an interacting DE, not just matter density ceases to evolve as a standard dust i.e. $a^{-3}(t)$ but its perturbations also experience  additional effects both from homogeneous (at background level) as well as the inhomogeneous (at perturbative level) DE. Similarly, the evolution of DE perturbations experience both homogeneous and inhomogeneous contribution from the matter sector. As a consequence, the evolution of cosmological perturbations becomes non-trivial. In fact, one observes following characteristic features in presence of an interacting DEM scenario : (i) enhancement of growth of matter perturbations, (ii) oscillations of DE perturbations about non-zero mean value, (iii) DE perturbations sources matter density perturbations, etc. These features  
being absent in the standard minimally coupled scalar field scenarios, such as quintessence, demands to re-perform a dynamical analysis under given notable effects.

Our dynamical analysis consists of an interacting DEM FRW background and linear matter density perturbations over it. Since the observable large scale structure and its evolution lies in the deep sub-horizon regime, therefore, for our analysis we consider effects of cosmological perturbations in that regime only. It is also beneficial in a sense that matter density perturbations becomes scale-independent which makes dynamical analysis independent of the perturbation scale. Hence, we have three dimensionless variables in total : two for the background which is constructed out of the kinetic and potential term of the scalar field in the Friedmann equation, and one for perturbations which consists of scale-independent matter density perturbations with an on-an-average effect of field perturbations. By simultaneously solving equations for our dimensionless quantities, we obtain fourteen critical points, out of which four are identical which give rise to same mathematical and physical description of our universe. Hence, in total we have ten distinct critical points which is ofcourse more than one would obtain in case of minimally coupled scenario.

We find that out of these ten distinct critical points only four is mathematically stable under linear perturbations in the solution space. However, from physical point of view only one point is stable i.e. CP(i) of table (\ref{tab:CP}), which is supposed to lie in the far future when field energy density gets completely dominated and growth of structure formation reduces to zero. Also, we find that all points could be saddle for a certain parametric regime of coupling parameter $n$, however, physically only one such point is possible i.e. CP(h) which only gives rise to a matter dominated description of the universe with a large growth of matter perturbations at high redshifts. Moreover, by simultaneously solving the set of autonomous equations, we find that the late-time behavior of dimensionless variables agrees with the stable CP(i). In particular, it also verify our previous found solution of $\vph$ in
\cite{MS-SS-IJMPD}, 
in which we have shown that this is the only real solution for a coupled set of background differential equations. 

Moreover, by formulating background dimensionless variables in terms of matter density parameter we have constructed a phase space solution between two observable $\Om$ and $f$. This we have done by fixing background dimensionless variables to that of the value corresponds to  stable CP(i). In that we find that only one physical trajectory is possible which leads to matter dominated to DE dominated era with decreasing $f$. However, a similar trajectory which also leads to the same configuration starts with negative $f$, which is not of physical importance. Further, in view of the necessity in the modifications in the analytical growth factor ansatz, we have also obtained a phase space dynamics of growth index $\c$ with $\Om$. In that we find that there lies only one such solution which is weakly dynamical. While all other solutions gets diverge in a finite interval of time, it correctly describe the high redshift approximation of $\c$ i.e. $6/11$. Also, a speed up evolution of $\c$ is observed near the present epoch till the asymptotic future.

\section*{Acknowledgments}

The work of MKS was supported by the Council of Scientific and Industrial Research (CSIR), Government of India.

\end{document}